\documentclass[preprint,showpacs,preprintnumbers,amsmath,amssymb]{revtex4}

\usepackage{graphicx}% Include figure files
\usepackage{dcolumn}% Align table columns on decimal point
\usepackage{bm}% bold math

\newcommand{\vecvar}[1]{\mbox{\boldmath$#1$}}
\newcommand{\vv}[1]{\mbox{\boldmath$#1$}}

\newcommand{\Tr}{{\rm Tr}}
\newcommand{\ee}{\begin{equation}\begin{aligned}}
\newcommand{\dd}{\end{aligned}\end{equation}}
\newcommand{\ar}[1]{{\left( #1 \right)}}
\newcommand{\m}{{N_{el}}}

\allowdisplaybreaks[1]

\begin{document}

%\preprint{APS/123-QED}

\title{Biorthogonal linear-scaling approach for the transcorrelated method}% Force line breaks with \\

\author{Masashi Kojo and Kikuji Hirose}
 \affiliation{Graduate School of Engineering, Osaka University, 2-1 Yamadaoka, Suita, Osaka 565-0871 Japan}

%\date{\today}

\begin{abstract}
We develop a biorthogonal linear-scaling algorithm for a transcorrelated method based on the localized nature of transformed orbitals. The transcorrelated method, which employs a similarity-transformed Hamiltonian referred to as a transcorrelated Hamiltonian, enables highly accurate first-principles condensed-matter calculations in principle. Meanwhile, the transcorrelated Hamiltonian numerically prevents us from applying it to large systems because the transcorrelated Hamiltonian is a non-Hermitian operator and contains a 3-body electron-electron interaction term. Non-Hermiticity means that left and right wave functions of the total energy expectation value are different from each other. Namely, a biorthogonal form is required. Our new method allows us to handle the non-Hermitian operator and exhibits a linear-scaling behavior.
\end{abstract}

\pacs{Valid PACS appear here}% PACS, the Physics and Astronomy
\maketitle

\section{Introduction}
  The transcorrelated method\cite{prsa01,prsa02,prsa03,prsa04,prsa05,jcp01,jcp03,jcp03a} , proposed by Boys and Handy, is regarded as one of the approaches based on the Jastrow-Slater-type wave function, such as the Variational Monte Carlo\cite{rmp01} method. The salient aspect of the transcorrelated method is its treatment of the Jastrow function. Instead of naively handling the Jastrow-Slater-type wave function, a similarity-transformed Hamiltonian with respect to the Jastrow function is used. Thanks to this particular Hamiltonian, referred to as a transcorrelated Hamiltonian, the many-body integral that stems from the function form of the Jastrow function is reduced to just a three-body one. Furthermore, we can obtain one-body energies that satisfy Koopmans theorem\cite{jcp03}. Considering that even within the single Slater determinant, the Jastrow-Slater-type wave function yields highly accurate ground-state energy, the transcorrelated method is thought to have great potential as a first-principles calculation tool. However, one must still tackle the non-Hermiticity of the transcorrelated Hamiltonian and the three-body electron-electron interaction term that is absent in the original Hamiltonian. The non-Hermiticity involves the biorthogonal treatment in which left and right eigenfunctions are different from each other. Furthermore, the three-body interaction term obviously limits the size of systems that we can investigate by the transcorrelated method, because it requires large computational cost that scales as $O \left( N_{el}^{3} \right)$. Here, $N_{el}$ denotes the number of electrons.

 The purpose of this paper is to propose an efficient $O \left( N_{el} \right)$ method of handling the above two issues with localized orbitals. In Sec. \ref{sec:001}, we introduce the transcorrelated method at first. And then we describe details of our approach in sec. \ref{sec:002}. In the first two subsections of Sec. \ref{sec:002}, we develop a biorthogonal self-consistent-field (SCF) equation for non-orthogonal orbitals. After that, we provide the important proof for handling the inverse of the overlapping matrix. Lastly, we discuss about the scaling behavior.

\section{Transcorrelated method}
\label{sec:001}
At first, we briefly introduce the biorthogonal transcorrelated method for the later discussion. In the transcorrelated method, instead of the ordinary many-body Hamiltonian $H$, one addresses the transcorrelated Hamiltonian 
\begin{equation}
\begin{aligned}
   H_{TC} = F^{-1} H F.
\end{aligned}
\end{equation}
Here, $F$ represents the Jastrow function. Because of the non-Hermiticity of the transcorrelated Hamiltonian $H_{TC}$, the variational approach to the energy expectation value of $H_{TC}$,
\begin{equation}
\label{eq:004}
   E = \int \Phi^\ast H_{TC} \Phi d^{N_{el}} \vecvar{r},
\end{equation}
 does not give the proper ground-state energy of the original Hamiltonian $H$, in principle. Instead, to obtain the proper ground state, the biorthogonal approach\cite{jcp01} must be taken, in which left and right wave functions are different from each other, as
\begin{equation}
\label{eq:005}
   E = \int \Phi_{L}^\ast H_{TC} \Phi_{R} d^{N_{el}} \vecvar{r}.
\end{equation}
In the formulation given below, the transcorrelated method in this biorthogonal form is discussed with the single Slater determinant. Spin indices are omitted for simplicity.

 Suppose, for instance, the function form of the Jastrow function is given by
\begin{equation}
  F = \exp \left[ -\frac{1}{2} \sum_{i=1}^{N_{el}} \sum_{j \neq i}^{N_{el}} u \left( \vecvar{r}_{i}, \vecvar{r}_{j} \right) \right],
\end{equation}
then the transcorrelated Hamiltonian can be explicitly written as\cite{jcp04,jcp03,jcp03a}
\begin{equation}
\label{eq:original_tc}
\begin{aligned}
  H_{TC} & \displaystyle = \sum_{i=1}^{N_{el}} \left[ -\frac{1}{2}\nabla_{i}^2 + v_{ext} \left( \vecvar{r}_i \right) \right] \\
  & + \frac{1}{2} \sum_{i=1}^{N_{el}} \sum_{j \neq i}^{N_{el}} v_{ee} \left( \vecvar{r}_{i}, \vecvar{r}_{j} \right) \\
  & + \frac{1}{6} \sum_{i=1}^{N_{el}} \sum_{j \neq i}^{N_{el}} \sum_{k \neq i,j}^{N_{el}} v_{eee} \left( \vecvar{r}_i, \vecvar{r}_j, \vecvar{r}_k \right )
\end{aligned}
\end{equation}
with the notations
\begin{equation}
\label{eq:002}
\begin{aligned}
  v_{ee} \left( \vecvar{r}_i, \vecvar{r}_j \right) & \displaystyle = \frac{1}{\left\vert \vecvar{r}_i - \vecvar{r}_j \right\vert} + \nabla_{i}^2 u \left( \vecvar{r}_i, \vecvar{r}_j \right) \\
  & - \nabla_i u \left( \vecvar{r}_i, \vecvar{r}_j \right) \cdot \nabla_i u \left( \vecvar{r}_i, \vecvar{r}_j \right) \\
  & + 2 \nabla_i u \left( \vecvar{r}_i, \vecvar{r}_j \right) \cdot \nabla_i,
\end{aligned}
\end{equation}
\begin{equation}
  v_{eee} \left( \vecvar{r}_i, \vecvar{r}_j, \vecvar{r}_k \right) = -3 \nabla_i u \left( \vecvar{r}_i, \vecvar{r}_j \right) \cdot \nabla_i u \left( \vecvar{r}_i, \vecvar{r}_k \right).
\end{equation}
%%In this case, the non-Hermitian terms are given by
%%\begin{equation}
%%\label{eq:006}
%%\begin{aligned}
%% & \frac{1}{2} \left( H_{TC} - H_{TC}^\dagger \right) \\
%% & = +\frac{1}{2} \sum_{i=1}^{N_{el}} \sum_{j \neq i}^{N_{el}} \left[ \nabla_{i}^2 u \left( \vecvar{r}_i, \vecvar{r}_j \right)+ 2 \nabla_i u \left( \vecvar{r}_i, \vecvar{r}_j \right) \cdot \nabla_i \right].
%%\end{aligned}
%%\end{equation}
Thus, the SCF equation and its Hamiltonian are written as
\begin{equation}
\label{eq:tc_scf_original}
\begin{aligned}
  H_{TC-SCF} \phi_{Ri} \left( \vecvar{x} \right) = \sum_{j=1}^{N_{el}} \varepsilon_{ji} \phi_{Rj} \left( \vecvar{x} \right)
\end{aligned}
\end{equation}
and
\begin{equation}
\label{eq:tc_hamil_original}
\begin{aligned}
  & H_{TC-SCF} \phi_{Ri} \left( \vecvar{x} \right)\\
  & = \left[ -\frac{1}{2}\nabla^{2} + v_{ext} \left( \vecvar{x} \right) \right] \phi_{Ri} \left( \vecvar{x} \right) \\
  & + \frac{1}{2} \sum_{j = 1}^{N_{el}} \int d \vecvar{y} \phi_{Lj}^\ast \left( \vecvar{y} \right) \left[ v_{ee}\left( \vecvar{x}, \vecvar{y} \right) + v_{ee} \left( \vecvar{y}, \vecvar{x} \right) \right] \\
  & \times \left\lVert \phi_{Ri}\left( \vecvar{x} \right) \phi_{Rj}\left( \vecvar{y} \right) \right\rVert \\
  & + \frac{1}{6} \sum_{j,k=1}^{N_{el}} \int d \vecvar{y} d \vecvar{z} \phi_{Lj}^\ast \left( \vecvar{y} \right) \phi_{Lk}^\ast \left( \vecvar{z} \right) \big[ v_{eee} \left( \vecvar{x}, \vecvar{y}, \vecvar{z} \right) \\
  & + v_{eee} \left( \vecvar{y}, \vecvar{z}, \vecvar{x} \right)+v_{eee} \left( \vecvar{z}, \vecvar{x}, \vecvar{y} \right) \big] \left\lVert \phi_{Ri}\left( \vecvar{x} \right) \phi_{Rj} \left( \vecvar{y} \right) \phi_{Rk} \left( \vecvar{z} \right) \right\rVert. \\
\end{aligned}
\end{equation}
Here, $\left\{ \varepsilon_{ij} \right\}$, $\left\{ \phi_{Li} \right\}$, $\left\{ \phi_{Ri} \right\}$, and $\left\lVert \cdots \right\rVert$ denote constants originating with the Lagrange multipliers, the left one-body wave functions, the right one-body wave functions, and the Slater determinant, respectively. It should be noted that the one-body wave functions $\left\{ \phi_{Li}, \phi_{Ri} \right\}$ are biorthogonal.

\section{Linear-scaling method in biorthogonal form}
\label{sec:002}
Here, we discuss a measure against the unfavorable scaling behavior on the number of electrons $N_{el}$. Utilizing the localized nature of transformed wave functions referred to as localized orbitals, the scaling behavior can be improved from $O \left( N_{el}^{3} \right)$ to $O \left( N_{el} \right)$. Yet the localized orbital is not a kind of orthonormal wave function in general; thus, Eq.(\ref{eq:tc_scf_original}) should be replaced with one without an orthogonality restriction. This idea is regarded as an extension of the work of Mauri et al.\cite{prb01,prb01a,pre01,pre01a} for linear scaling calculations.

\subsection{Energy expression without orthogonality restriction}
\label{subsec:001}

 Let us consider, for instance, the two-body interaction term
\begin{equation}
\begin{aligned}
 & \Tr \left[ v_{ee} \left( \vecvar{x}, \vecvar{x}^\prime \right) \rho \left( \vecvar{x}, \vecvar{x}^\prime; \vecvar{y}, \vecvar{y}^\prime \right) \right],
\end{aligned}
\end{equation}
where
\begin{equation}
\label{eq:st_two_body}
\begin{aligned}
 & \rho \left( \vecvar{x}, \vecvar{x}^\prime; \vecvar{y}, \vecvar{y}^\prime \right) = \frac{N_{el} \left( N_{el}-1 \right)}{2} \int d \vecvar{x}_3 \cdots d \vecvar{x}_{N_{el}} \\
 & \times F^{-1} \Psi \left( \vecvar{x}, \vecvar{x}^\prime, \vecvar{x}_3, \cdots, \vecvar{x}_{N_{el}} \right) \Psi^\ast \left( \vecvar{y}, \vecvar{y}^\prime, \vecvar{x}_3, \cdots, \vecvar{x}_{N_{el}} \right) F \\
\end{aligned}
\end{equation}
with $\Psi$ being the exact many-body wave function of the ordinary many-body Hamiltonian $H$. The two-body reduced density matrix Eq.(\ref{eq:st_two_body}) has a different definition compared with the ordinary one \cite{book01} due to the existence of the Jastrow function $F$. Therefore, here and hereafter, we call it the similarity-transformed two-body reduced matrix. As discussed in Appendix, since left and right eigenvectors of the transcorrelated Hamiltonian $H_{TC}$, $\left\{ \Psi F, F^{-1} \Psi \right\}$, in Eq.(\ref{eq:st_two_body}) are approximately treated as Slater determinants in the transcorrelated method, the similarity-transformed one-body reduced density matrix,
\begin{equation}
\label{eq:st_one_body}
\begin{aligned}
 & \rho \left( \vecvar{x}; \vecvar{y} \right) = N_{el} \int d \vecvar{x}_2 \cdots d \vecvar{x}_{N_{el}}  \\
 & \times F^{-1} \Psi \left( \vecvar{x}, \vecvar{x}_2, \cdots, \vecvar{x}_{N_{el}} \right) \Psi^\ast \left( \vecvar{y}, \vecvar{x}_2, \cdots, \vecvar{x}_{N_{el}} \right) F, \\
\end{aligned}
\end{equation}
satisfies the idempotency condition as in the case of Hartree-Fock method, and the cumulant expansion is also applicable to the similarity-transformed two-body reduced density matrix Eq.(\ref{eq:st_two_body}) as follows :
\begin{equation}
\label{eq:cumulant}
\begin{aligned}
 \rho \left( \vecvar{x}, \vecvar{x}^\prime; \vecvar{y}, \vecvar{y}^\prime \right)
 & = \frac{1}{2} \rho \left( \vecvar{x}; \vecvar{y} \right) \rho \left( \vecvar{x}^\prime; \vecvar{y}^\prime \right) - \frac{1}{2} \rho \left( \vecvar{x}^\prime; \vecvar{y} \right) \rho \left( \vecvar{x}; \vecvar{y}^\prime \right).
\end{aligned}
\end{equation}
Also, according to the proof in Appendix, the similarity-transformed one-body reduced density matrix is expressed in terms of its left and right eigenvectors, $\left\{ \psi_{Li} \right\}$ and $\left\{ \psi_{Ri} \right\}$, as
\begin{equation}
\label{eq:1bodyrdm}
\begin{aligned}
 \rho \left( \vecvar{x}; \vecvar{y} \right) = \sum_{i=1}^{N_{el}} \psi_{Ri}\left( \vecvar{x} \right) \psi_{Li}^\ast \left( \vecvar{y} \right).
\end{aligned}
\end{equation}
%Here $\left\{ \psi_{Li} \right\}$ and $\left\{ \psi_{Ri} \right\}$ denote left and right eigenvectors of the one-body reduced density matrix, respectively (see the reference \cite{book01} and Appendix for more details) .
%Namely, eigenvalues are $1$ for occupied orbitals and $0$ for unoccupied orbitals. Therefore, Eq.(\ref{eq:1bodyrdm}) is rewritten as
%the similarity transformation in Eq.(\ref{eq:st_one_body}) conserves the eigenvalues of the one-body reduced density matrix and the one-body reduced density matrix must satisfy the idempotency condition, eigenvalues are $1$ for occupied orbitals and $0$ for unoccupied orbitals. Therefore, Eq.(\ref{eq:1bodyrdm}) is rewritten as
%\begin{equation}
%\label{eq:rho}
%\begin{aligned}
% \rho \left( \vecvar{x}; \vecvar{y} \right) = \sum_{i=1}^{N_{el}} \psi_{Ri}\left( \vecvar{x} \right) \psi_{Li}^\ast \left( \vecvar{y} \right).
%\end{aligned}
%\end{equation}
%It should be noted that the applicability of the cumulant expansion in Eq.(\ref{eq:cumulant}) also derives from this approximation.

Consider the non-biorthonormal orbitals $\left\{ \phi_{Li} \right\}$ and $\left\{ \phi_{Ri} \right\}$ defined as
\begin{equation}
\label{eq:nonono}
\begin{aligned}
 \phi_{Li} \left( \vecvar{x} \right) = \sum_{j=1}^{N_{el}} \omega_{i j} \psi_{Lj} \left( \vecvar{x} \right), \\
 \phi_{Ri} \left( \vecvar{x} \right) = \sum_{j=1}^{N_{el}} \omega_{i j} \psi_{Rj} \left( \vecvar{x} \right).
\end{aligned}
\end{equation}
Conversely, from Eq.(\ref{eq:nonono}), one obtains
\begin{equation}
\label{eq:nononoinv}
\begin{aligned}
 \psi_{Li} \left( \vecvar{x} \right) = \sum_{j=1}^{N_{el}} \omega_{i j}^{-1} \phi_{Lj} \left( \vecvar{x} \right), \\
 \psi_{Ri} \left( \vecvar{x} \right) = \sum_{j=1}^{N_{el}} \omega_{i j}^{-1} \phi_{Rj} \left( \vecvar{x} \right).
\end{aligned}
\end{equation}
Now, the overlapping matrix between $\left\{ \phi_{Li} \right\}$ and $\left\{ \phi_{Ri} \right\}$ is 
\begin{equation}
\begin{aligned}
 S_{jk} & = \int d \vecvar{x} \phi_{Lj}^\ast \left( \vecvar{x} \right) \phi_{Rk} \left( \vecvar{x} \right) \\
        & = \sum_{i,i^\prime=1}^{N_{el}} \omega_{j i}^\ast \omega_{k i^\prime} \int d \vecvar{x} \psi_{L i}^\ast \left( \vecvar{x} \right) \psi_{R i^\prime} \left( \vecvar{x} \right) \\
        & = \sum_{i=1}^{N_{el}} \omega_{j i}^\ast \omega_{k i}.\\
\end{aligned}
\end{equation}
Therefore, the inverse of the overlapping matrix is
\begin{equation}
\label{eq:s_inv}
\begin{aligned}
 S_{jk}^{-1} & = \sum_{i=1}^{N_{el}} \omega_{i j}^{-1} \omega_{i k}^{-\ast}.\\
\end{aligned}
\end{equation}
Note that $\omega_{i k}^{-\ast}$ means $\left( \omega^{-1} \right)_{i k}^{\ast}$. Substituting Eqs.(\ref{eq:nononoinv}) and (\ref{eq:s_inv}) into Eq.(\ref{eq:1bodyrdm}) yields
\begin{equation}
\begin{aligned}
 \rho \left( \vecvar{x}; \vecvar{y} \right) & = \sum_{i=1}^{N_{el}} \psi_{R i}\left( \vecvar{x} \right) \psi_{L i}^\ast \left( \vecvar{y} \right) \\
 & = \sum_{i=1}^{N_{el}} \left[ \sum_{j=1}^{N_{el}} \omega_{ij}^{-1} \phi_{Rj}\left( \vecvar{x} \right) \right] \left[ \sum_{k=1}^{N_{el}} \omega_{ik}^{-\ast} \phi_{Lk}^\ast \left( \vecvar{y} \right) \right]\\
 & = \sum_{k=1}^{N_{el}} \sum_{j=1}^{N_{el}} \left( \sum_{i=1}^{N_{el}} \omega_{i j}^{-1} \omega_{i k}^{-\ast} \right) \phi_{R j}\left( \vecvar{x} \right) \phi_{L k}^\ast \left( \vecvar{y} \right) \\
 & = \sum_{k=1}^{N_{el}} \sum_{j=1}^{N_{el}} S_{jk}^{-1} \phi_{R j}\left( \vecvar{x} \right) \phi_{L k}^\ast \left( \vecvar{y} \right).\\
\end{aligned}
\end{equation}
As a consequence, the two-body interaction term becomes
\begin{equation}
\begin{aligned}
 & \Tr \left[ v_{ee} \left( \vecvar{x}, \vecvar{x}^\prime \right) \rho \left( \vecvar{x}, \vecvar{x}^\prime; \vecvar{y}, \vecvar{y}^\prime \right) \right] \\
 & = \int d \vecvar{x} d \vecvar{x}^\prime d \vecvar{y} d \vecvar{y}^\prime \delta\left( \vecvar{x}-\vecvar{y} \right) \delta\left( \vecvar{x}^\prime-\vecvar{y}^\prime \right) v_{ee} \left( \vecvar{x}, \vecvar{x}^\prime \right) \\
 & \times \left[ \frac{1}{2} \rho \left( \vecvar{x}; \vecvar{y} \right) \rho \left( \vecvar{x}^\prime; \vecvar{y}^\prime \right)-\frac{1}{2} \rho \left( \vecvar{x}^\prime; \vecvar{y} \right) \rho \left( \vecvar{x}; \vecvar{y}^\prime \right) \right] \\
 & = \int d \vecvar{x} d \vecvar{x}^\prime d \vecvar{y} d \vecvar{y}^\prime \delta\left( \vecvar{x}-\vecvar{y} \right) \delta\left( \vecvar{x}^\prime-\vecvar{y}^\prime \right) v_{ee} \left( \vecvar{x}, \vecvar{x}^\prime \right) \\
 & \times \frac{1}{2} \sum_{i,j,k,l=1}^{N_{el}} S_{ij}^{-1} S_{kl}^{-1} \big[ \phi_{Ri}\left( \vecvar{x} \right) \phi_{Lj}^\ast \left( \vecvar{y} \right) \phi_{Rk}\left( \vecvar{x}^\prime \right) \phi_{Ll}^\ast \left( \vecvar{y}^\prime \right) \\
 & - \phi_{Ri}\left( \vecvar{x}^\prime \right) \phi_{Lj}^\ast \left( \vecvar{y} \right) \phi_{Rk}\left( \vecvar{x} \right) \phi_{Ll}^\ast \left( \vecvar{y}^\prime \right) \big] \\
 & = \frac{1}{2} \sum_{i,j,k,l=1}^{N_{el}} S_{ij}^{-1} S_{kl}^{-1} \int d \vecvar{x} d \vecvar{x}^\prime \phi_{Lj}^\ast \left( \vecvar{x} \right)\phi_{Ll}^\ast \left( \vecvar{x}^\prime \right) v_{ee} \left( \vecvar{x}, \vecvar{x}^\prime \right) \\
 & \times \left\lVert \phi_{Ri}\left( \vecvar{x} \right) \phi_{Rk}\left( \vecvar{x}^\prime \right) \right\rVert. \\
\end{aligned}
\end{equation}
In a similar fashion, since the three-body reduced density matrix can be decomposed as
\begin{equation}
\begin{aligned}
 & \rho \left( \vecvar{x}, \vecvar{x}^\prime, \vecvar{x}^{\prime \prime} ; \vecvar{y}, \vecvar{y}^\prime, \vecvar{y}^{\prime \prime} \right) \\
 & = \frac{1}{6} \begin{vmatrix}
 \rho \left( \vecvar{x}; \vecvar{y} \right) & \rho \left( \vecvar{x}; \vecvar{y}^\prime \right) & \rho \left( \vecvar{x}; \vecvar{y}^{\prime \prime} \right) \\
 \rho \left( \vecvar{x}^\prime; \vecvar{y} \right) & \rho \left( \vecvar{x}^\prime; \vecvar{y}^\prime \right) & \rho \left( \vecvar{x}^\prime; \vecvar{y}^{\prime \prime} \right) \\
 \rho \left( \vecvar{x}^{\prime \prime}; \vecvar{y} \right) & \rho \left( \vecvar{x}^{\prime \prime}; \vecvar{y}^\prime \right) & \rho \left( \vecvar{x}^{\prime \prime}; \vecvar{y}^{\prime \prime} \right) \\
  \end{vmatrix}, \\
\end{aligned}
\end{equation}
the total energy without the orthogonality restriction is expressed by
\begin{equation}
\label{eq:total_energy}
\begin{aligned}
 E & = \Tr \Big\{ \sum_{i=1}^{N_{el}} \left[ -\frac{1}{2}\nabla_{i}^{2} + v_{ext} \left( \vecvar{x}_{i} \right) \right] \rho\left( \vecvar{x}_{i}; \vecvar{y}_{i} \right) \\
   & + \sum_{i,j=1}^{N_{el}} v_{ee}\left( \vecvar{x}_{i}, \vecvar{x}_{j} \right) \rho \left( \vecvar{x}_{i}, \vecvar{x}_{j}; \vecvar{y}_{i}, \vecvar{y}_{j} \right) \\
   & + \sum_{i,j,k=1}^{N_{el}} v_{eee} \left( \vecvar{x}_{i}, \vecvar{x}_{j}, \vecvar{x}_{k} \right) \rho \left( \vecvar{x}_{i}, \vecvar{x}_{j}, \vecvar{x}_{k}; \vecvar{y}_{i}, \vecvar{y}_{j}, \vecvar{y}_{k} \right) \Big\} \\
 & = \sum_{i.j=1}^{N_{el}} S_{ij}^{-1} \int d \vecvar{x} \phi_{Lj}^\ast \left( \vecvar{x} \right) \left[ -\frac{1}{2} \nabla^2 + v_{ext}\left( \vecvar{x} \right) \right] \phi_{Ri}\left( \vecvar{x} \right) \\
 & + \frac{1}{2} \sum_{i,j,k,l=1}^{N_{el}} S_{ij}^{-1} S_{kl}^{-1} \int d \vecvar{x} d \vecvar{x}^\prime \phi_{Lj}^\ast \left( \vecvar{x} \right)\phi_{Ll}^\ast \left( \vecvar{x}^\prime \right) v_{ee} \left( \vecvar{x}, \vecvar{x}^\prime \right) \\
 & \times \left\lVert \phi_{Ri}\left( \vecvar{x} \right) \phi_{Rk}\left( \vecvar{x}^\prime \right) \right\rVert\\
 & + \frac{1}{6} \sum_{i,j,k,l,m,n=1}^{N_{el}} S_{ij}^{-1} S_{kl}^{-1} S_{mn}^{-1} \int d \vecvar{x} d \vecvar{x}^\prime d \vecvar{x}^{\prime \prime} \phi_{Lj}^\ast \left( \vecvar{x} \right) \phi_{Ll}^\ast \left( \vecvar{x}^\prime \right) \\
 & \times \phi_{Ln}^\ast \left( \vecvar{x}^{\prime\prime} \right) v_{eee} \left( \vecvar{x}, \vecvar{x}^\prime, \vecvar{x}^{\prime \prime} \right) \left\lVert \phi_{Ri}\left( \vecvar{x} \right) \phi_{Rk} \left( \vecvar{x}^\prime \right) \phi_{Rm}\left( \vecvar{x}^{\prime\prime} \right)\right\rVert. \\
\end{aligned}
\end{equation}

\subsection{Biorthogonal SCF equation without orthogonality constraint}
\label{subsec:002}

 The biorthogonal SCF equation is derived by the variational approach to the total energy $E$ in Eq.(\ref{eq:total_energy}), namely, $\partial E / \partial \phi_{Lp}^\ast \left( \vecvar{w} \right) = 0$ and $ \partial E / \partial \phi_{Rp}^\ast \left( \vecvar{w} \right) = 0$. For instance, the functional derivative of the two-body term $E_{2}$ in the total energy $E$ with respect to $\phi_{Lp}^\ast \left( \vecvar{w} \right)$ is 
\begin{widetext}
\begin{equation}
\begin{aligned}
 \frac{\partial E_{2}}{\partial \phi_{Lp}^\ast \left( \vecvar{w} \right)} & = \frac{1}{2} \sum_{i,j,k,l=1}^{N_{el}} \frac{\partial S_{ij}^{-1}}{\partial \phi_{Lp}^\ast \left( \vecvar{w} \right)} S_{kl}^{-1} \int d \vecvar{x} d \vecvar{y} \phi_{Lj}^\ast \left( \vecvar{x} \right) \phi_{Ll}^\ast \left( \vecvar{y} \right) v_{ee} \left( \vecvar{x}, \vecvar{y} \right) \left\lVert \phi_{Ri}\left( \vecvar{x} \right) \phi_{Rk} \left( \vecvar{y} \right) \right\rVert \\
 & + \frac{1}{2} \sum_{i,j,k,l=1}^{N_{el}} S_{ij}^{-1} \frac{\partial S_{kl}^{-1}}{\partial \phi_{Lp}^\ast \left( \vecvar{w} \right)} \int d \vecvar{x} d \vecvar{y} \phi_{Lj}^\ast \left( \vecvar{x} \right) \phi_{Ll}^\ast \left( \vecvar{y} \right) v_{ee} \left( \vecvar{x}, \vecvar{y} \right) \left\lVert \phi_{Ri}\left( \vecvar{x} \right) \phi_{Rk} \left( \vecvar{y} \right) \right\rVert \\
 & + \frac{1}{2} \sum_{i,j,k,l=1}^{N_{el}} S_{ij}^{-1} S_{kl}^{-1} \int d \vecvar{x} d \vecvar{y} \delta_{jp}\delta\left( \vecvar{x}-\vecvar{w} \right) \phi_{Ll}^\ast \left( \vecvar{y} \right) v_{ee} \left( \vecvar{x}, \vecvar{y} \right) \left\lVert \phi_{Ri}\left( \vecvar{x} \right) \phi_{Rk} \left( \vecvar{y} \right) \right\rVert \\
 & + \frac{1}{2} \sum_{i,j,k,l=1}^{N_{el}} S_{ij}^{-1} S_{kl}^{-1} \int d \vecvar{x} d \vecvar{y} \phi_{Lj}^\ast \left( \vecvar{x} \right) \delta_{lp}\delta\left( \vecvar{y}-\vecvar{w} \right) v_{ee} \left( \vecvar{x}, \vecvar{y} \right) \left\lVert \phi_{Ri}\left( \vecvar{x} \right) \phi_{Rk} \left( \vecvar{y} \right) \right\rVert \\
 & = \sum_{i,j=1}^{N_{el}} \frac{\partial S_{ij}^{-1}}{\partial \phi_{Lp}^\ast \left( \vecvar{w} \right)} \int d \vecvar{x} \phi_{Lj}^\ast \left( \vecvar{x} \right) \bigg\{ \frac{1}{2} \sum_{k,l=1}^{N_{el}} S_{kl}^{-1} \int d \vecvar{y} \phi_{Ll}^\ast \left( \vecvar{y} \right) \left[ v_{ee} \left( \vecvar{x}, \vecvar{y} \right) + v_{ee} \left( \vecvar{y}, \vecvar{x} \right) \right] \\
 & \times \left\lVert \phi_{Ri}\left( \vecvar{x} \right) \phi_{Rk} \left( \vecvar{y} \right) \right\rVert \bigg\} \\
 & + \sum_{i=1}^{N_{el}} S_{ip}^{-1} \left\{ \frac{1}{2} \sum_{k,l=1}^{N_{el}} S_{kl}^{-1} \int d \vecvar{y} \phi_{Ll}^\ast \left( \vecvar{y} \right) \left[ v_{ee} \left( \vecvar{w}, \vecvar{y} \right) + v_{ee} \left( \vecvar{y}, \vecvar{w} \right) \right] \left\lVert \phi_{Ri}\left( \vecvar{w} \right) \phi_{Rk} \left( \vecvar{y} \right) \right\rVert \right\} \\
 & = \sum_{i,j=1}^{N_{el}} \frac{\partial S_{ij}^{-1}}{\partial \phi_{Lp}^\ast \left( \vecvar{w} \right)} \int d \vecvar{x} \phi_{Lj}^\ast \left( \vecvar{x} \right) H_{TC-SCF2}^{R} \phi_{Ri}\left( \vecvar{x} \right) + \sum_{i=1}^{N_{el}} S_{ip}^{-1} H_{TC-SCF2}^{R} \phi_{Ri}\left( \vecvar{w} \right). \\
\end{aligned}
\end{equation}
Here, $H_{TC-SCF2}^{R}$ is the operator defined as 
\begin{equation}
\begin{aligned}
  H_{TC-SCF2}^{R} \phi_{Ri}\left( \vecvar{x} \right) = \frac{1}{2} \sum_{k,l=1}^{N_{el}} S_{kl}^{-1} \int d \vecvar{y} \phi_{Ll}^\ast \left( \vecvar{y} \right) \left[ v_{ee} \left( \vecvar{x}, \vecvar{y} \right) + v_{ee} \left( \vecvar{y}, \vecvar{x} \right) \right] \left\lVert \phi_{Ri}\left( \vecvar{x} \right) \phi_{Rk} \left( \vecvar{y} \right) \right\rVert.
\end{aligned}
\end{equation}
In exactly the same way, one readily obtains
\begin{equation}
\label{eq:e2_tc_derived}
\begin{aligned}
  \frac{\partial E}{\partial \phi_{Lp}^\ast \left( \vecvar{w} \right)} & = \sum_{i,j=1}^{N_{el}} \frac{\partial S_{ij}^{-1}}{\partial \phi_{Lp}^\ast \left( \vecvar{w} \right)} \int d \vecvar{x} \phi_{Lj}^\ast \left( \vecvar{x} \right) H_{TC-SCF}^{R} \phi_{Ri}\left( \vecvar{x} \right) + \sum_{i=1}^{N_{el}} S_{ip}^{-1} H_{TC-SCF}^{R} \phi_{Ri}\left( \vecvar{w} \right),
\end{aligned}
\end{equation}
where
\begin{equation}
\label{eq:tc_scf_unconstrained}
\begin{aligned}
 & H_{TC-SCF}^{R} \phi_{Ri} \left( \vecvar{x} \right) \\
 & = \left[ -\frac{1}{2}\nabla^{2} + v_{ext} \left( \vecvar{x} \right) \right] \phi_{Ri} \left( \vecvar{x} \right) + \frac{1}{2} \sum_{j,k=1}^{N_{el}} S_{jk}^{-1} \int d \vecvar{y} \phi_{Lk}^\ast \left( \vecvar{y} \right) \left[ v_{ee}\left( \vecvar{x}, \vecvar{y} \right) + v_{ee} \left( \vecvar{y}, \vecvar{x} \right) \right] \left\lVert \phi_{Ri}\left( \vecvar{x} \right) \phi_{Rj}\left( \vecvar{y} \right) \right\rVert \\
  & + \frac{1}{6} \sum_{j,k,l,m=1}^{N_{el}} S_{jk}^{-1} S_{lm}^{-1} \int d \vecvar{y} d \vecvar{z} \phi_{Lk}^\ast \left( \vecvar{y} \right) \phi_{Lm}^\ast \left( \vecvar{z} \right) \big[ v_{eee} \left( \vecvar{x}, \vecvar{y}, \vecvar{z} \right) + v_{eee} \left( \vecvar{y}, \vecvar{z}, \vecvar{x} \right)+v_{eee} \left( \vecvar{z}, \vecvar{x}, \vecvar{y} \right) \big] \\
 & \times \left\lVert \phi_{Ri}\left( \vecvar{x} \right) \phi_{Rj} \left( \vecvar{y} \right) \phi_{Rl} \left( \vecvar{z} \right) \right\rVert. \\
\end{aligned}
\end{equation}
The functional derivative of the inverse matrix $S^{-1}$ with respect to $\phi_{Lp}^\ast \left( \vecvar{w} \right)$ in Eq.(\ref{eq:e2_tc_derived}) is not straightforwardly computable so we therefore compute it by utilizing the relationship $S S^{-1} = I$ as,
\begin{equation}
\begin{aligned}
 \frac{\partial S_{ij}^{-1}}{\partial \phi_{Lp}^\ast \left( \vecvar{w} \right)} & = - \left[ S^{-1} \frac{\partial S}{\partial \phi_{Lp}^\ast \left( \vecvar{w} \right) } S^{-1} \right]_{ij} = - \sum_{kl}^{N_{el}} S_{ik}^{-1} \delta_{kp} \phi_{Rl}\left( \vecvar{w} \right) S_{lj}^{-1} = - \sum_{l}^{N_{el}} S_{ip}^{-1} S_{lj}^{-1} \phi_{Rl}\left( \vecvar{w} \right). 
\end{aligned}
\end{equation}
Substituting the above-mentioned result into Eq.(\ref{eq:e2_tc_derived}) yields
\begin{equation}
\begin{aligned}
 \frac{\partial E}{\partial \phi_{Lp}^\ast \left( \vecvar{w} \right)} & = - \sum_{i,j,l=1}^{N_{el}} S_{ip}^{-1} S_{lj}^{-1} \phi_{Rl}\left( \vecvar{w} \right) \int d \vecvar{x} \phi_{Lj}^\ast \left( \vecvar{x} \right) H_{TC-SCF}^{R} \phi_{Ri}\left( \vecvar{x} \right) + \sum_{i=1}^{N_{el}} S_{ip}^{-1} H_{TC-SCF}^{R} \phi_{Ri}\left( \vecvar{w} \right). \\
\end{aligned}
\end{equation}
Also, operating $\sum_{p=1}^{N_{el}} S_{pk}$ from its left side, one has
\begin{equation}
\begin{aligned}
 \sum_{p=1}^{N_{el}} S_{pk} \frac{\partial E}{\partial \phi_{Lp}^\ast \left( \vecvar{w} \right)} & = - \sum_{l=1}^{N_{el}} \left\{ \sum_{j=1}^{N_{el}} S_{lj}^{-1} \int d \vecvar{x} \phi_{Lj}^\ast \left( \vecvar{x} \right) H_{TC-SCF}^{R} \phi_{Rk}\left( \vecvar{x} \right) \right\} \phi_{Rl}\left( \vecvar{w} \right) + H_{TC-SCF}^{R} \phi_{Rk}\left( \vecvar{w} \right) \\
 & = - \sum_{l=1}^{N_{el}} \varepsilon_{Rlk} \phi_{Rl}\left( \vecvar{w} \right) + H_{TC-SCF}^{R} \phi_{Rk}\left( \vecvar{w} \right),
\end{aligned}
\end{equation}
where
\begin{equation}
\label{eq:tc_scf_equation_unconstrained_energy}
\begin{aligned}
  \varepsilon_{Rji} = \sum_{k=1}^{N_{el}} S_{jk}^{-1} \int d \vecvar{y} \phi_{Lk}^\ast \left( \vecvar{y} \right) H_{TC-SCF}^{R} \phi_{Ri} \left( \vecvar{y} \right).
\end{aligned}
\end{equation}
Because $\partial E/\partial \phi_{Lp}^\ast \left( \vecvar{w} \right) = 0$, the SCF equation is finally derived as
\begin{equation}
\label{eq:tc_scf_equation_unconstrained}
\begin{aligned}
  H_{TC-SCF}^{R} \phi_{Ri} \left( \vecvar{x} \right) = \sum_{j=1}^{N_{el}} \varepsilon_{Rji} \phi_{Rj} \left( \vecvar{x} \right).
\end{aligned}
\end{equation}

Similarly, $H_{TC-SCF}^{L}$, the SCF equation, and $\varepsilon_{Lji}$ corresponding to $\left\{ \phi_{Li} \right\}$ are obtained as Eqs.(\ref{eq:tc_scf_unconstrained}), (\ref{eq:tc_scf_equation_unconstrained}), and (\ref{eq:tc_scf_equation_unconstrained_energy}), respectively.
\end{widetext}

\subsection{Handling $S^{-1}$}
\label{subsec:003}

The inverse of the overlapping matrix, $S^{-1}$, requires $O \left( N_{el}^{3} \right)$ computation. It should therefore be replaced with the polynomial function of $S$\cite{prb01,prb01a} , namely,
\begin{equation}
\label{eq:q}
\begin{aligned}
 S^{-1} \backsimeq Q = \sum_{n=0}^{N_{S}} \left( I - S \right)^{n}
\end{aligned}
\end{equation}
with an odd number $N_{S}$. As discussed later, it is straightforward to show that the computational cost for performing Eq.(\ref{eq:tc_scf_equation_unconstrained}) is reduced to $O \left( N_{el} \right)$ by utilizing localized orbitals accompanied with Eq.(\ref{eq:q}).

Here, we discuss whether the replacement (\ref{eq:q}) causes any errors in the total energy. To do so, we decompose the overlapping matrix $S$ with its eigenvalue matrix $\Lambda$, left eigenvector matrix $V$, and right eigenvector matrix $U$ as
\begin{equation}
\begin{aligned}
  S = U \Lambda V.
\end{aligned}
\end{equation}
It is easily shown, by using $\Theta = \sum_{n=0}^{N_{S}} \left( I-\Lambda \right)^{n}$ and $\Xi = \left( I-\Lambda \right)^{N_{S}+1} - I$, that the eigenvalue matrix of $ Q-S^{-1} $ is reduced to $- \Lambda^{-1} \left( I+\Xi \right)$ as
\begin{equation}
\label{eq:q_invs_is_negative}
\begin{aligned}
  V \left( Q-S^{-1} \right) U & = - \Lambda^{-1} + \Theta \\
  & = - \Lambda^{-1} \left[ I - \Lambda \Theta \right] \\
  & = - \Lambda^{-1} \left[ I + \left( I-\Lambda \right) \Theta - \Theta \right] \\
  & = - \Lambda^{-1} \left[ I + \sum_{n=1}^{N_{S}+1} \left( I-\Lambda \right)^{n} - \Theta \right] \\
  & = - \Lambda^{-1} \left[ I + \left( I-\Lambda \right)^{N_{S}+1} -  I \right] \\
  & = - \Lambda^{-1} \left( I+\Xi \right). \\
\end{aligned}
\end{equation}
The non-negative definiteness of $S$ makes $\Lambda^{-1}$ a non-negative definite matrix. Also, $I+\Xi = \left( I-\Lambda \right)^{N_{S}+1}$ is a non-negative definite matrix for odd number $N_{S}$. The matrix $ Q-S^{-1} $ is therefore a non-positive definite matrix because both $\Lambda^{-1}$ and $I+\Xi$ are non-negative definite matrices.  Using $\Lambda$, $U$, and $V$, one obtains
\begin{equation}
\begin{aligned}
  Q_{ij}Q_{kl} & = \left( \sum_{p=1}^{N_{el}} U_{ip} \Theta_{pp} V_{pj} \right) \left( \sum_{q=1}^{N_{el}} U_{kq} \Theta_{qq} V_{ql} \right) \\
               & = \sum_{p,q=1}^{N_{el}} U_{ip} U_{kq} \Theta_{pp} \Theta_{qq} V_{ql} V_{pk}.
\end{aligned}
\end{equation}
This is a spectrum expansion of the $N_{el}^{2} \times N_{el}^{2}$ matrix ${}^{2}Q$ which has an element $Q_{ij}Q_{kl}$ in the $\left(i,k\right)$-th row and the $\left( j,l\right)$-th column. Thus, $\Theta_{pp} \Theta_{qq}$ is the $\left( p,q \right)$-th eigenvalue because of
\begin{equation}
\begin{aligned}
 \sum_{j,l=1}^{N_{el}} Q_{ij}Q_{kl} U_{jp} U_{lq} & = \sum_{j=1}^{N_{el}} Q_{ij} U_{jp} \sum_{l=1}^{N_{el}} Q_{kl} U_{lq} \\
 & = \Theta_{pp} \Theta_{qq} U_{ip} U_{kq}. \\
\end{aligned}
\end{equation}
If $N_{el}^{2} \times N_{el}^{2}$ matrix ${}^{2}S^{-1}$ is defined as
\begin{equation}
  {}^{2}S^{-1} \equiv \left( S_{ij}^{-1} S_{kl}^{-1} \right)_{N_{el}^2 \times N_{el}^2},
\end{equation} 
${}^{2}Q-{}^{2}S^{-1}$ is proved to be a non-positive definite matrix as follows:
\begin{equation}
\begin{aligned}
  & \Theta_{pp} \Theta_{qq} - \Lambda_{pp}^{-1} \Lambda_{qq}^{-1}\\
  & = -\Lambda_{pp}^{-1} \Lambda_{qq}^{-1} \left( 1 - \Lambda_{pp} \Lambda_{qq} \Theta_{pp} \Theta_{qq} \right) \\
  & = -\Lambda_{pp}^{-1} \Lambda_{qq}^{-1} \left[ 1 - \left( - \Lambda \Theta \right)_{pp} \left( - \Lambda \Theta \right)_{qq} \right] \\
  & = -\Lambda_{pp}^{-1} \Lambda_{qq}^{-1} \left( 1 - \Xi_{pp} \Xi_{qq} \right) \\
  & \leq 0
\end{aligned}
\end{equation}
within the assumption of $0 \leq \Lambda_{pp} \leq 2$. By the same token, if $N_{el}^{3} \times N_{el}^{3}$ matrices ${}^{3}Q$ and ${}^{3}S^{-1}$ are defined as
\begin{equation}
  {}^{3}Q \equiv \left( Q_{ij}Q_{kl}Q_{mn} \right)_{N_{el}^3 \times N_{el}^3},
\end{equation}
and
\begin{equation}
  {}^{3}S^{-1} \equiv \left( S_{ij}^{-1} S_{kl}^{-1} S_{mn}^{-1} \right)_{N_{el}^3 \times N_{el}^3},
\end{equation}
respectively, ${}^{3}Q-{}^{3}S^{-1}$ is also proved to be a non-positive definite matrix as
\begin{equation}
\begin{aligned}
  & \Theta_{pp} \Theta_{qq} \Theta_{rr} - \Lambda_{pp}^{-1} \Lambda_{qq}^{-1} \Lambda_{rr}^{-1}\\
  & = -\Lambda_{pp}^{-1} \Lambda_{qq}^{-1} \Lambda_{rr}^{-1} \left( 1 - \Lambda_{pp} \Lambda_{qq} \Lambda_{rr} \Theta_{pp} \Theta_{qq} \Theta_{rr} \right) \\
  & = -\Lambda_{pp}^{-1} \Lambda_{qq}^{-1} \Lambda_{rr}^{-1} \left( 1 + \Xi_{pp} \Xi_{qq} \Xi_{rr} \right) \\
  & \leq 0.
\end{aligned}
\end{equation}
As a result, according to the proof by Mauri et al.\cite{prb01,prb01a} , it can be proved that even when $S^{-1}$ is replaced with $Q$, one obtains the same total energy $E$ by restricting $\Lambda_{pp}$ in the range $\left[ 0, 2 \right]$. The restriction of $\Lambda_{pp}$ in this range is easily achieved by adjusting the trace of $S$.

\subsection{Scaling behavior}
\label{subsec:004}

Now, we discuss the scaling behavior of the above-mentioned scheme. For example, consider the exchange term in the total energy replaced with $Q$,
\begin{equation}
\begin{aligned}
  & \frac{1}{2} \sum_{i,j,k,l=1}^{N_{el}} Q_{ij} Q_{kl} \int d \vecvar{x} d \vecvar{y} \phi_{Lj}^\ast \left( \vecvar{x} \right) \phi_{Ll}^\ast \left( \vecvar{y} \right) v_{ee} \left( \vecvar{x}, \vecvar{y} \right) \\
  & \times \phi_{Ri}\left( \vecvar{y} \right) \phi_{Rk}\left( \vecvar{x} \right). \\
\end{aligned}
\end{equation}
The term $Q_{ij}$ means that the index $j$ runs over the adjacent indices to the index $i$. Similarly, $Q_{kl}$ and $ \phi_{Ll}^\ast \left( \vecvar{y} \right) \phi_{Ri}\left( \vecvar{y} \right) $ restricts the indices $k$ and $l$ on the adjacent indices to the index $i$. Ultimately, the only index $i$ runs over all indices. Thus, this exchange term shows an $O \left( N_{el} \right)$ behavior accompanied with localized orbitals.

Next, consider the three-body interaction terms. Classifying by the number of interchanged pairs, there are three cases, namely, for example,
\ee
 E_3^{Non} & = \frac{1}{6} \sum_{i,j,k,l,m,n=1}^{N_{el}} Q_{ij}Q_{kl}Q_{mn} \int d \vv x d \vv x' d \vv x'' \phi^*_{Lj}( \vv x )\phi^*_{Ll}( \vv x' ) \\
 & \times \phi^*_{Ln}( \vv x'') v_{eee}(\vv x, \vv x', \vv x'') \phi_{Ri}(\vv x)\phi_{Rk}(\vv x')\phi_{Rm}(\vv x''),
\dd
\ee
 E_3^{One} & = \frac{1}{6} \sum_{i,j,k,l,m,n=1}^{N_{el}} Q_{ij}Q_{kl}Q_{mn} \int d \vv x d \vv x' d \vv x'' \phi^*_{Lj}( \vv x )\phi^*_{Ll}( \vv x' ) \\
 & \times \phi^*_{Ln}( \vv x'') v_{eee}(\vv x, \vv x', \vv x'') \phi_{Ri}(\vv x)\phi_{Rm}(\vv x')\phi_{Rk}(\vv x''),
\dd
\ee
 E_3^{Two} & = \frac{1}{6} \sum_{i,j,k,l,m,n=1}^{N_{el}} Q_{ij}Q_{kl}Q_{mn} \int d \vv x d \vv x' d \vv x'' \phi^*_{Lj}( \vv x )\phi^*_{Ll}( \vv x' ) \\
 & \times \phi^*_{Ln}( \vv x'') v_{eee}(\vv x, \vv x', \vv x'') \phi_{Rm}(\vv x)\phi_{Ri}(\vv x')\phi_{Rk}(\vv x'').
\dd
In the non-interchanged case, $E_3^{Non}$, one obtains
\ee
 E_3^{Non} & = \frac{1}{6} \sum_{i,j,k,l,m,n=1}^{N_{el}} Q_{ij}Q_{kl}Q_{mn} \int d \vv x d \vv x' d \vv x'' \phi^*_{Lj}( \vv x )\phi^*_{Ll}( \vv x' ) \\
 & \times \phi^*_{Ln}( \vv x'') v_{eee}(\vv x, \vv x', \vv x'') \phi_{Ri}(\vv x)\phi_{Rk}(\vv x')\phi_{Rm}(\vv x'') \\
 & = \frac{1}{6} \int d \vv x \left\{\sum_{i,j=1}^{N_{el}} Q_{ij} \phi^*_{Lj}( \vv x )\phi_{Ri}(\vv x)\right\} \int d \vv x' \left\{ \sum_{k,l=1}^\m Q_{kl} \phi^*_{Ll}( \vv x' )\phi_{Rk}(\vv x') \right\} \\
 & \times  \int d \vv x'' v_{eee}(\vv x, \vv x', \vv x'') \left\{ \sum_{m,n=1}^\m Q_{mn} \phi^*_{Ln}( \vv x'') \phi_{Rm}(\vv x'') \right\} \\
 & = \frac{1}{6} \int d \vv x \rho( \vv x; \vv x) \int d \vv x' \rho( \vv x'; x') \int d \vv x'' v_{eee}(\vv x, \vv x', \vv x'') \rho ( \vv x''; \vv x''). \\
\dd
The cost which we pay for this is just an $O \ar{\m}$ to calculate $\rho( \vv x; \vv x)$. Therefore, it is computable with an $O\ar{\m}$ cost. Similarly, the one-pair interchanged case, $E_3^{One}$, can be rewritten as
\ee
 E_3^{One} & = \frac{1}{6} \sum_{i,j,k,l,m,n=1}^{N_{el}} Q_{ij}Q_{kl}Q_{mn} \int d \vv x d \vv x' d \vv x'' \phi^*_{Lj}( \vv x )\phi^*_{Ll}( \vv x' ) \\
 & \times \phi^*_{Ln}( \vv x'') v_{eee}(\vv x, \vv x', \vv x'') \phi_{Ri}(\vv x)\phi_{Rm}(\vv x')\phi_{Rk}(\vv x'')\\
 & = \frac{1}{6} \int d \vv x \left\{\sum_{i,j=1}^\m Q_{ij} \phi^*_{Lj}( \vv x )\phi_{Ri}(\vv x) \right\} \sum_{k,l,m,n=1}^\m Q_{kl}Q_{mn} \int d \vv x' d \vv x'' \phi^*_{Ll}( \vv x' ) \\
 & \times \phi^*_{Ln}( \vv x'') v_{eee}(\vv x, \vv x', \vv x'') \phi_{Rm}(\vv x')\phi_{Rk}(\vv x'')\\
 & = \frac{1}{6} \int d \vv x \rho \ar{ x; x } \sum_{k,l,m,n=1}^\m Q_{kl}Q_{mn} \int d \vv x' d \vv x'' \phi^*_{Ll}( \vv x' ) \\
 & \times \phi^*_{Ln}( \vv x'') v_{eee}(\vv x, \vv x', \vv x'') \phi_{Rm}(\vv x')\phi_{Rk}(\vv x'').
\dd
As in the case of the exchange term, only one index among $k$,$l$,$m$, and $n$ runs over all indices. Thus, it also shows a linear-scaling behavior. Note that $\rho \ar{ x; x }$ can be computed separately and beforehand. In the case of the two-pair interchanged case, $E_3^{Two}$, the terms $Q_{ij}$, $Q_{kl}$, $Q_{mn}$, $\phi^*_{Lj}( \vv x )\phi_{Rm}(\vv x)$, $\phi^*_{Ll}( \vv x' )\phi_{Ri}(\vv x')$, and $\phi^*_{Ln}( \vv x'')\phi_{Rk}(\vv x'')$ restrict 5 indices. Thus, only one index among 6 indices runs over all indices. As a result, the three-body interaction terms also show a linear-scaling behavior. Likewise, the other terms which are not mentioned here in the total energy and the SCF equation are computable in $O \left( N_{el} \right)$ computational costs.\\

\section{Conclusion}
\label{sec:003}

 In summary, we have proposed an efficient approach for the transcorrelated method in the biorthogonal form. In this approach, the original transcorrelated Hamiltonian is replaced with the one free of the orthogonality constraint in order to handle the non-orthogonal localized orbitals. Because of being free of the orthogonality restriction, the unhandy inverse matrix $S^{-1}$ appears in the formulation. To achieve a linear-scaling behavior of the whole framework, we replaced $S^{-1}$ with the polynomial function of $S$ referred to as $Q$, and we gave the proof that even when $S^{-1}$ is replaced with $Q$, the same ground state is obtainable. Lastly, we showed that our approach has an $O \left( N_{el} \right)$ scaling behavior.

 For now, in this paper, we proposed and discussed the theoretical framework only. To prove the effectivity of our approach, a number of numerical examples are required indeed. Currently we are under investigation in this direction.

%It is though to enable us to handle the larger systems with the transcorrelated method.

\begin{acknowledgments}
 We wish to acknowledge support through the Global COE gCenter of Excellence for Atomically Controlled Fabrication Technologyh, and Scientific Research in Priority Areas gDevelopment of New Quantum Simulators and Quantum Designh (Grant No. 17064012) from the Ministry of Education, Culture, Sports, Science and Technology.
\end{acknowledgments}
\appendix

\section{Cumulant expansion of the similarity-transformed density matrix}

Consider the exact many-body wave function $\Psi$ which satisfies the shr\"odinger equation
\ee
 H \Psi = E \Psi.
\dd
Then the exact left and right eigenvectors of the transcorrelated Hamiltonian $H_{TC} \ar{ = F^{-1} H F }$ are written with this $\Psi$ as
\ee
 \Psi_L = \Psi F,
\dd
\ee
 \Psi_R = F^{-1} \Psi.
\dd
These two eigenvectors are approximately treated as Slater determinants in the transcorrelated method, namely,
\ee
 \Psi_L = \Vert \psi_{L1} \psi_{L2} \cdots \psi_{L\m} \Vert,
\dd
\ee
 \Psi_R = \Vert \psi_{R1} \psi_{R2} \cdots \psi_{R\m} \Vert.
\dd
We are not approximating the exact many-body wave function $\Psi$ solely. We have a high degree of freedom for $\{ \psi_{Li}, \psi_{Ri}  \}$ as in the case of the Hartree-Fock method, because Slater determinants are invariant under any unitary transformations. So we can select biorthogonal sets as $\{ \psi_{Li}, \psi_{Ri}  \}$. From the things mentioned above, the similarity-transformed many-body density matrix is derived as \\
\ee
\label{stmbdm}
 & \rho \ar{\vv x_1, \vv x_2, \cdots, \vv x_\m; \vv y_1, \vv y_2, \cdots, \vv y_\m} \\
 & = \Psi_R \ar{\vv x_1, \vv x_2, \cdots, \vv x_\m} \Psi_L^\ast \ar{ \vv y_1, \vv y_2, \cdots, \vv y_\m} \\
 & = \frac{1}{\m!} \begin{vmatrix}
      \psi_{R1} \ar{\vv  x_1} & \psi_{R2} \ar{\vv  x_1} & \cdots & \psi_{R\m} \ar{\vv x_1} \\
      \psi_{R1} \ar{\vv  x_2} & \psi_{R2} \ar{\vv  x_2} & \cdots & \psi_{R\m} \ar{\vv x_2} \\
      \vdots               & \vdots               &        & \vdots \\
      \psi_{R1} \ar{\vv x_\m} & \psi_{R2} \ar{\vv x_\m} & \cdots & \psi_{R\m} \ar{\vv x_\m}\\
     \end{vmatrix}
     \begin{vmatrix}
      \psi_{L1} \ar{\vv y_1}  & \psi_{L2} \ar{\vv y_1}  & \cdots & \psi_{L\m} \ar{\vv y_1} \\
      \psi_{L1} \ar{\vv y_2}  & \psi_{L2} \ar{\vv y_2}  & \cdots & \psi_{L\m} \ar{\vv y_2} \\
      \vdots               & \vdots               &        & \vdots \\
      \psi_{L1} \ar{\vv y_\m} & \psi_{L2} \ar{\vv y_\m} & \cdots & \psi_{L\m} \ar{\vv y_\m}\\
     \end{vmatrix}^{\dagger} \\
 & = \frac{1}{\m!} \begin{vmatrix}
      \rho \ar{\vv  x_1; \vv y_1} & \rho \ar{\vv  x_1; \vv y_2} & \cdots & \rho \ar{\vv x_1; \vv y_\m} \\
      \rho \ar{\vv  x_2; \vv y_1} & \rho \ar{\vv  x_2; \vv y_2} & \cdots & \rho \ar{\vv x_2; \vv y_\m} \\
      \vdots                      & \vdots                      &        & \vdots \\
      \rho \ar{\vv x_\m; \vv y_1} & \rho \ar{\vv x_\m; \vv y_2} & \cdots & \rho \ar{\vv x_\m; \vv y_\m}\\
     \end{vmatrix}. \\
\dd
Here
\ee
\rho(\vv x; \vv y) = \sum_{i=1}^{N_{el}} \psi_{Ri} (\vv x) \psi_{Li}^\ast (\vv y).
\dd
This is exactly the same definition with that of the reference \cite{book01} except the definition of the one-body reduced density matrix. This difference between definitions of the one-body one leaves no effect on the applicability of the cumulant expansion when the one-body one is still idempotent. In other words, if the one-body one satisfies
\ee
\int d \vv y \rho \ar{\vv x; \vv y} \rho\ar{\vv y; \vv x'} = \rho\ar{\vv x; \vv x'}, 
\dd
the cumulant expansion is also available to the similarity-transformed ones. It is readily shown that the biorthogonality of $\{ \psi_{Li}, \psi_{Ri} \}$ proves an idempotence of the one-body reduced density matrix. 

\subsubsection{Proof of the availability of the cumulant expansion to the similarity-transformed reduced density matrix}
 At first, we define a similarity-transformed $M$-body reduced density matrix as
\ee
 &  \rho \ar{\vv x_1, \vv x_2, \cdots, \vv x_M; \vv y_1, \vv y_2, \cdots, \vv y_M} \\
 & = \binom{\m}{M} \int \Psi_R \ar{\vv x_1, \vv x_2, \cdots, \vv x_\m} \Psi_L^\ast \ar{\vv y_1, \vv y_2, \cdots, \vv y_\m} \prod_{k=M+1}^{\m} d \vv x_k d \vv y_k \delta \ar{\vv x_k-\vv y_k}.
\dd
From this definition, a $\m$-body one is
\ee
\label{m_body_one}
 &   \rho\ar{\vv x_1, \vv x_2, \cdots, \vv x_\m; \vv y_1, \vv y_2, \cdots, \vv y_\m} \\
 & = \binom{\m}{\m} \int \Psi_R \ar{\vv x_1, \vv x_2, \cdots, \vv x_\m} \Psi_L^\ast \ar{\vv y_1, \vv y_2, \cdots, \vv y_\m} \prod_{k=\m+1}^{\m} d \vv x_k d \vv y_k \delta \ar{\vv x_k-\vv y_k} \\
 & = \Psi_R \ar{\vv x_1, \vv x_2, \cdots, \vv x_\m} \Psi_L^\ast \ar{\vv y_1, \vv y_2, \cdots, \vv y_\m} \\
 & = \frac{1}{\m!} \begin{vmatrix}
      \rho \ar{\vv  x_1; \vv y_1} & \rho \ar{\vv  x_1; \vv y_2} & \cdots & \rho \ar{\vv x_1; \vv y_\m} \\
      \rho \ar{\vv  x_2; \vv y_1} & \rho \ar{\vv  x_2; \vv y_2} & \cdots & \rho \ar{\vv x_2; \vv y_\m} \\
      \vdots                      & \vdots                      &        & \vdots \\
      \rho \ar{\vv x_\m; \vv y_1} & \rho \ar{\vv x_\m; \vv y_2} & \cdots & \rho \ar{\vv x_\m; \vv y_\m}\\
     \end{vmatrix}, \\
\dd
as shown in Eq.(\ref{stmbdm}). Also it is easily checked that a $\ar{M-1}$-body one can be obtained from a $M$-body one as
\ee
  & \rho\ar{\vv x_1, \vv x_2, \cdots, \vv x_{M-1}; \vv y_1, \vv y_2, \cdots, \vv y_{M-1}} \\
  & = \frac{\binom{\m}{M-1}}{\binom{\m}{M}} \int d \vv x_M d \vv y_M \delta \ar{\vv x_M-\vv y_M} \rho\ar{\vv x_1, \vv x_2, \cdots, \vv x_M; \vv y_1, \vv y_2, \cdots, \vv y_M} \\
  & = \frac{M}{\m-M+1} \int d \vv x_M d \vv y_M \delta \ar{\vv x_M-\vv y_M} \rho\ar{\vv x_1, \vv x_2, \cdots, \vv x_M; \vv y_1, \vv y_2, \cdots, \vv y_M}. \\
\dd
When the $M$-body one is given as
\ee
 &   \rho\ar{\vv x_1, \vv x_2, \cdots, \vv x_M; \vv y_1, \vv y_2, \cdots, \vv y_M} \\
 & = \frac{1}{M!} \begin{vmatrix}
      \rho \ar{\vv x_1; \vv y_1} & \rho \ar{\vv x_1; \vv y_2} & \cdots & \rho \ar{\vv x_1; \vv y_M} \\
      \rho \ar{\vv x_2; \vv y_1} & \rho \ar{\vv x_2; \vv y_2} & \cdots & \rho \ar{\vv x_2; \vv y_M} \\
      \vdots                     & \vdots                     &        & \vdots \\
      \rho \ar{\vv x_M; \vv y_1} & \rho \ar{\vv x_M; \vv y_2} & \cdots & \rho \ar{\vv x_M; \vv y_M}\\
     \end{vmatrix}, \\
\dd
the $\ar{M-1}$-body one is
{\footnotesize
\begin{align*}
  & \rho\ar{\vv x_1, \vv x_2, \cdots, \vv x_{M-1}; \vv y_1, \vv y_2, \cdots, \vv y_{M-1}} \\
  & = \frac{M}{\m-M+1} \int d \vv x_M d \vv y_M \delta \ar{\vv x_M-\vv y_M} \rho\ar{\vv x_1, \vv x_2, \cdots, \vv x_M; \vv y_1, \vv y_2, \cdots, \vv y_M} \\
  & = \frac{\ar{\m-M+1}^{-1}}{\ar{M-1}!} \int d \vv x_M d \vv y_M \delta \ar{\vv x_M-\vv y_M} \begin{vmatrix}
      \rho \ar{\vv x_1; \vv y_1} & \rho \ar{\vv x_1; \vv y_2} & \cdots & \rho \ar{\vv x_1; \vv y_M} \\
      \rho \ar{\vv x_2; \vv y_1} & \rho \ar{\vv x_2; \vv y_2} & \cdots & \rho \ar{\vv x_2; \vv y_M} \\
      \vdots                     & \vdots                     &        & \vdots \\
      \rho \ar{\vv x_M; \vv y_1} & \rho \ar{\vv x_M; \vv y_2} & \cdots & \rho \ar{\vv x_M; \vv y_M}\\
     \end{vmatrix} \\
  & = \frac{\ar{\m-M+1}^{-1}}{\ar{M-1}!} \int d \vv x_M \begin{vmatrix}
      \rho \ar{\vv x_1; \vv y_1} & \rho \ar{\vv x_1; \vv y_2} & \cdots & \rho \ar{\vv x_1; \vv x_M} \\
      \rho \ar{\vv x_2; \vv y_1} & \rho \ar{\vv x_2; \vv y_2} & \cdots & \rho \ar{\vv x_2; \vv x_M} \\
      \vdots                     & \vdots                     &        & \vdots \\
      \rho \ar{\vv x_M; \vv y_1} & \rho \ar{\vv x_M; \vv y_2} & \cdots & \rho \ar{\vv x_M; \vv x_M}\\
     \end{vmatrix} \\
  & = \frac{\ar{\m-M+1}^{-1}}{\ar{M-1}!} \int d \vv x_M \bigg\{ \rho \ar{\vv x_M; \vv y_M} \begin{vmatrix}
      \rho \ar{\vv x_1; \vv y_1} & \rho \ar{\vv x_1; \vv y_2} & \cdots & \rho \ar{\vv x_1; \vv y_{M-1}} \\
      \rho \ar{\vv x_2; \vv y_1} & \rho \ar{\vv x_2; \vv y_2} & \cdots & \rho \ar{\vv x_2; \vv y_{M-1}} \\
      \vdots                     & \vdots                     &        & \vdots                     \\
      \rho \ar{\vv x_{M-1}; \vv y_1} & \rho \ar{\vv x_{M-1}; \vv y_2} & \cdots & \rho \ar{\vv x_{M-1}; \vv y_{M-1}} \\
     \end{vmatrix} \\
  & +\sum_{k=1}^{M-1} \ar{-1}^{M+k} \rho \ar{\vv x_M; \vv y_k} \begin{vmatrix}
      \rho \ar{\vv x_1; \vv y_1} & \cdots & \rho \ar{\vv x_1; \vv y_{k-1}} & \rho \ar{\vv x_1; \vv y_{k+1}} & \cdots & \rho \ar{\vv x_1; \vv x_M} \\
      \rho \ar{\vv x_2; \vv y_1} & \cdots & \rho \ar{\vv x_2; \vv y_{k-1}} & \rho \ar{\vv x_2; \vv y_{k+1}} & \cdots & \rho \ar{\vv x_2; \vv x_M} \\
      \vdots                     & \vdots &                                & \vdots                         &        & \vdots                     \\
      \rho \ar{\vv x_{M-1}; \vv y_1} & \cdots & \rho \ar{\vv x_{M-1}; \vv y_{k-1}} & \rho \ar{\vv x_{M-1}; \vv y_{k+1}} & \cdots & \rho \ar{\vv x_{M-1}; \vv x_M}\\
     \end{vmatrix} \bigg\} \\
  & = \frac{\ar{\m-M+1}^{-1}}{\ar{M-1}!} \bigg\{ \int d \vv x_M \rho \ar{\vv x_M; \vv y_M} \begin{vmatrix}
      \rho \ar{\vv x_1; \vv y_1} & \rho \ar{\vv x_1; \vv y_2} & \cdots & \rho \ar{\vv x_1; \vv y_{M-1}} \\
      \rho \ar{\vv x_2; \vv y_1} & \rho \ar{\vv x_2; \vv y_2} & \cdots & \rho \ar{\vv x_2; \vv y_{M-1}} \\
      \vdots                     & \vdots                     &        & \vdots                     \\
      \rho \ar{\vv x_{M-1}; \vv y_1} & \rho \ar{\vv x_{M-1}; \vv y_2} & \cdots & \rho \ar{\vv x_{M-1}; \vv y_{M-1}} \\
     \end{vmatrix} \\
  & +\sum_{k=1}^{M-1} \ar{-1}^{M+k} \begin{vmatrix}
      \rho \ar{\vv x_1; \vv y_1} & \cdots & \rho \ar{\vv x_1; \vv y_{k-1}} & \rho \ar{\vv x_1; \vv y_{k+1}} & \cdots & \int d \vv x_M \rho \ar{\vv x_1; \vv x_M}\rho \ar{\vv x_M; \vv y_k}  \\
      \rho \ar{\vv x_2; \vv y_1} & \cdots & \rho \ar{\vv x_2; \vv y_{k-1}} & \rho \ar{\vv x_2; \vv y_{k+1}} & \cdots & \int d \vv x_M\rho \ar{\vv x_2; \vv x_M}\rho \ar{\vv x_M; \vv y_k} \\
      \vdots                     & \vdots                     &        & \vdots                     &        & \vdots                     \\
      \rho \ar{\vv x_{M-1}; \vv y_1} & \cdots & \rho \ar{\vv x_{M-1}; \vv y_{k-1}} & \rho \ar{\vv x_{M-1}; \vv y_{k+1}} & \cdots & \int d \vv x_M \rho \ar{\vv x_{M-1}; \vv x_M}\rho \ar{\vv x_M; \vv y_k} \\
     \end{vmatrix} \bigg\} \\
  & = \frac{\ar{\m-M+1}^{-1}}{\ar{M-1}!} \bigg\{ \m \begin{vmatrix}
      \rho \ar{\vv x_1; \vv y_1} & \rho \ar{\vv x_1; \vv y_2} & \cdots & \rho \ar{\vv x_1; \vv y_{M-1}} \\
      \rho \ar{\vv x_2; \vv y_1} & \rho \ar{\vv x_2; \vv y_2} & \cdots & \rho \ar{\vv x_2; \vv y_{M-1}} \\
      \vdots                     & \vdots                     &        & \vdots                     \\
      \rho \ar{\vv x_{M-1}; \vv y_1} & \rho \ar{\vv x_{M-1}; \vv y_2} & \cdots & \rho \ar{\vv x_{M-1}; \vv y_{M-1}} \\
     \end{vmatrix} \\
  & +\sum_{k=1}^{M-1} \ar{-1}^{M+k} \ar{-1}^{M-k-1} \begin{vmatrix}
      \rho \ar{\vv x_1; \vv y_1} & \rho \ar{\vv x_1; \vv y_2} & \cdots & \rho \ar{\vv x_1; \vv y_{M-1}} \\
      \rho \ar{\vv x_2; \vv y_1} & \rho \ar{\vv x_2; \vv y_2} & \cdots & \rho \ar{\vv x_2; \vv y_{M-1}} \\
      \vdots                     & \vdots                     &        & \vdots                     \\
      \rho \ar{\vv x_{M-1}; \vv y_1} & \rho \ar{\vv x_{M-1}; \vv y_2} & \cdots & \rho \ar{\vv x_{M-1}; \vv y_{M-1}} \\
     \end{vmatrix} \bigg\} \\
  & = \frac{\ar{\m-M+1}^{-1}}{\ar{M-1}!} \ar{\m-M+1} \begin{vmatrix}
      \rho \ar{\vv x_1; \vv y_1} & \rho \ar{\vv x_1; \vv y_2} & \cdots & \rho \ar{\vv x_1; \vv y_{M-1}} \\
      \rho \ar{\vv x_2; \vv y_1} & \rho \ar{\vv x_2; \vv y_2} & \cdots & \rho \ar{\vv x_2; \vv y_{M-1}} \\
      \vdots                     & \vdots                     &        & \vdots                     \\
      \rho \ar{\vv x_{M-1}; \vv y_1} & \rho \ar{\vv x_{M-1}; \vv y_2} & \cdots & \rho \ar{\vv x_{M-1}; \vv y_{M-1}} \\
     \end{vmatrix} \\
  & = \frac{1}{\ar{M-1}!} \begin{vmatrix}
      \rho \ar{\vv x_1; \vv y_1} & \rho \ar{\vv x_1; \vv y_2} & \cdots & \rho \ar{\vv x_1; \vv y_{M-1}} \\
      \rho \ar{\vv x_2; \vv y_1} & \rho \ar{\vv x_2; \vv y_2} & \cdots & \rho \ar{\vv x_2; \vv y_{M-1}} \\
      \vdots                     & \vdots                     &        & \vdots                     \\
      \rho \ar{\vv x_{M-1}; \vv y_1} & \rho \ar{\vv x_{M-1}; \vv y_2} & \cdots & \rho \ar{\vv x_{M-1}; \vv y_{M-1}}
     \end{vmatrix}.
\end{align*}}
As a consequence, by the mathematical induction, if the $\m$-body density matrix is given as in Eq.(\ref{stmbdm}) and the one-body reduced density matrix is idempotent, then the $M$-body reduced density matrix for an arbitrary $M \in [1,\m]$ is expressed as
\ee
 &   \rho\ar{\vv x_1, \vv x_2, \cdots, \vv x_M; \vv y_1, \vv y_2, \cdots, \vv y_M} \\
 & = \frac{1}{M!} \begin{vmatrix}
      \rho \ar{\vv x_1; \vv y_1} & \rho \ar{\vv x_1; \vv y_2} & \cdots & \rho \ar{\vv x_1; \vv y_M} \\
      \rho \ar{\vv x_2; \vv y_1} & \rho \ar{\vv x_2; \vv y_2} & \cdots & \rho \ar{\vv x_2; \vv y_M} \\
      \vdots                     & \vdots                     &        & \vdots \\
      \rho \ar{\vv x_M; \vv y_1} & \rho \ar{\vv x_M; \vv y_2} & \cdots & \rho \ar{\vv x_M; \vv y_M}\\
     \end{vmatrix}. \\
\dd

%\newpage %Just because of unusual number of tables stacked at end
\bibliography{bibdata}% Produces the bibliography via BibTeX.

\end{document}